\documentstyle[12pt,epsfig]{article}
\title{True parameters of $^4_{\Sigma}$He hypernucleus}
\author{V.M.Kolybasov \thanks{e--mail: kolybasv@sci.lebedev.ru}
 \\ Lebedev Physical Institute, 117924 Moscow, Russia}
\date{}
\begin{document}
\maketitle
\begin{abstract}
\normalsize
\baselineskip=24pt
It is shown that the true parameters of $^4_{\Sigma}$He differ from
the observed ones. The reason is that the amplitude of $^4_{\Sigma}$He
production in the reaction $^4$He(K$^-,{\pi}^-$) sharply varies just in
the corresponding mass region. It leads to the small, but noticeable
shift of the binding energy and the width. Besides, the data at the
threshold of ${\Sigma}^0$ production is found to give an additional
evidence that $^4_{\Sigma}$He width cannot exceed $8.0 \div 8.5$ MeV.
\\[\baselineskip]
\end{abstract}

\newpage
\baselineskip=24pt

The recent study of the missing mass spectrum from the reaction
\begin{equation}
^4{\mbox{He}}({\mbox K}^-,{\pi}^-)
\end{equation}
at 600 MeV/c has revealed a peak in the region close to $\Sigma$
production [1]. The peak corresponds to the bound $^4_{\Sigma}$He
state with parameters $E_{ex}=-7$ MeV and $\Gamma=7$ MeV ($E_{ex}$ is
the missing mass to a pion measured from the sum of masses ${\Sigma}^0
+ ^3$He). Here we give the central values of the parameters, obtained
in Ref.\ [1] with the help of several simple versions of the background
approximation.\footnote{The threshold of $({\Sigma}^- + {\mbox t})$
channel is situated 2.6 MeV lower than the threshold of $({\Sigma}^0 +
^3{\mbox{He}})$ channel. So $E_{ex} =-7$ MeV corresponds to the binding
energy 4.4 MeV.}

A production of $^4_{\Sigma}$He actually realizes the unique case when
the amplitude of resonance production is sharply varying function of
mass just in the region of resonance mass. If the amplitude strongly
changes on the resonance width, it can lead to the appreciable shift of
the observed energy and width compared with the true values. As will be
shown below, the amplitude of $^4_{\Sigma}$He production in the process
(1) rapidly varies in the $E_{ex}$ interval from $-10$ MeV to 0, that is,
just in the region of $^4_{\Sigma}$He mass.

The missing mass spectrum from the reaction
\begin{equation}
^4{\mbox{He}}({\mbox K}^-,{\pi}^+)
\end{equation}
was also studied in Ref.\ [1]. A comparison of the data on channels (1)
and (2) shows the following. Almost all events of the reaction (1) are
situated at $E_{ex}>0$. They form a broad maximum which is usually
associated with a quasifree ${\Sigma}^-$ production [2]. In contrast to
this, the channel (2) has also many events in the $E_{ex}$ region from
$-40$ to $-10$ MeV. They are obviously due to the tail of $\Lambda$
production. Besides, the channel (2) has an enhancement at $E_{ex}$ near
zero, and a distinct peak at $E_{ex}=-7$ MeV corresponding to the bound
$^4_{\Sigma}$He state (see the histogram in Fig.\ 1). In what follows,
primary attention will be focused on the region of the resonance peak.
However at first it is necessary to discuss a physical nature of the
``background'', that is the tail of $\Lambda$ production and the region of
``quasifree'' $\Sigma$ production. An attempt to describe $E_{ex}$ spectrum
from $-40$ to $-20$ MeV in terms of quasifree $\Lambda$ production was
unsuccessful. The corresponding curve decreased too rapidly in obvious
contradiction with the data. The model of quasifree $\Lambda$ production
with subsequent rescattering on residual nuclear system therefore was
applied. $^4$He wave function in oscillator potential was used, and
the oscillator parameter $p_0$ was considered as fitting one. Small values
of $p_0$ result in too rapidly decreasing curve. For large $p_0$ values
the curve, on the contrary, falls down too slowly, that contradicts the
data at $E_{ex}$ in the region $30 \div 50$ MeV. The optimum value is
$p_0=150$ MeV/c which leads to the curve in Fig.\ 1. Certainly, such
procedure is rather rough, and the corresponding errors will be considered
later on.

As to the region $E_{ex}>0$, the simplest mechanisms are quasifree $\Sigma$
production which can be accompanied by $\Sigma$ rescattering on residual
nuclear system or by ${\Sigma} \to {\Lambda}$ conversion. The analysis of
$({\mbox K}^-,{\pi}^{\pm})$ reactions on $^9$Be, $^{12}$C and $^4$He was
performed in Ref.\ [3]. It shows that the quasifree production gives a peak
which is narrower and leftward shifted compared with the experimental one.
At the same time an account of elastic and inelastic rescatterings together
with the interference of corresponding amplitudes results in a good
description of $E_{ex}$ spectra. Thus, it is possible to suppose that the
origin of a main part of the Fig.\ 1 spectrum  in general is clear. It
allows to study the region of $^4_{\Sigma}$He peak and $E_{ex}\sim 0$ in
more detail.

The most probable mechanism of $^4_{\Sigma}$He production is presented in
Fig.\ 2 (a) and (b). At first, $\Sigma$--hyperon is born on one of
neutrons, and then it coalesce with residual nuclear system. The difference
between Fig.\ 2 (a) and (b) is that in the first case we have the
two--particle intermediate state $^3$He -- ${\Sigma}^0$, and in the second
case the three--particle state p -- d -- ${\Sigma}^0$ is present (there
could also be four--particle state p -- p -- n -- ${\Sigma}^0$). An
amplitude of Fig.\ 2 (a) has singularities of two kinds: the root threshold
singularity at $E_{ex}=0$, and the triangle logarithmic singularity
located in complex plane. The latter is also situated near $E_{ex}=0$ for
kinematical conditions of Ref.\ [1]. Modulus squared of the triangle graph
amplitude $M_{\triangle}$ for this case is shown (without Breit--Wigner
factor) by solid curve in Fig.~3. Here and further we use the oscillator
wave function of $^4$He with the parameter $p_0=90$ MeV/c which
gives the best description of $^4$He(e,ep) data at small spectator momenta
[4]. According to the evaluation of Ref.\ [1], the bound state of
$^4_{\Sigma}$He is located at $E_{ex}=-7$ MeV and has the width about
7 MeV. Figure 3 shows that $|M_{\triangle}|^2$ strongly varies on the
resonance width. It can noticeably influence the result of $^4_{\Sigma}$He
parameters estimation from the experimental data.

It is worth noting that such sharp behavior of the amplitude is well
known for the case of stopped kaon capture ${\mbox K}^- {\mbox d} \to
{\mbox p} {\Lambda} {\pi}^-$. The pion spectrum for this reaction has a
distinct peak [5] associated with the triangle graph with the conversion
${\Sigma} \to {\Lambda}$ (see Ref.\ [6]). A cusp behavior was also
indicated in Ref.\ [7] devoted to stopped K$^-$ capture in $^4$He.
A possibility of the distortion for Breit--Wigner form in the case of
nearthreshold resonance was also mentioned in Ref.\ [8].

Sharp behavior of $|M_{\triangle}|^2$ is characteristic only for the
triangle graph of Fig.\ 2 (a) with a two--particle intermediate state.
The graph of Fig.\ 2 (b) with a three--particle intermediate state leads
to the smooth amplitude whose maximum is shifted rightward (see also
Ref.\ [9]). It is shown by dotted curve in Fig.\ 3. Therefore the
comparative contribution of Fig.\ 2 (a) and (b) graphs is rather
important. It is determined by the relations of the left lower and upper
vertices of both graphs. As to the nuclear vertices, their relation can
be obtained from the available data on $^4$He(e,ep) reaction [4]. They
show that the vertex of two--particle $^4$He decay is much more than the
vertices of three and four--particle decays at relative momenta up to
250 MeV/c. There are no direct information on the vertices of virtual
$^4_{\Sigma}$He  decays. Owing to a lack of data on sigma--nuclear
interactions, the reliable evaluations are now hardly possible. However,
we have no reasons to assume that the two--particle channel is preferred
here. Therefore it is possible to assert only that the contribution of
Fig.\ 2 (a) graph in any case should be noticeable against a
``background'' of Fig.\ 2 (b) graph. It is indirectly confirmed also by
the results of Ref.\ [7].

There is one more interesting point originating during the analysis of
the data. Modulus squared of Fig.\ 2 (a) graph is the product of the
modulus squared of triangle graph with constant lower vertex, which is
shown by solid line in Fig.\ 3, and Breit--Wigner resonance factor. It
is easy to see that this product has two peaks. The first corresponds to
the resonance, and the second is located at $E_{ex} \approx 0$. A ratio
of these peaks depends on the width of the resonance. For the case of
narrow resonance, the ``cusp'' maximum near $E_{ex}=0$ will be suppressed
by Breit--Wigner factor, and for the case of broad resonance this
maximum will be large. It imposes additional constraints on the resonance
width. The point is that, though the data have an enhancement near
$E_{ex}=0$, it is not large. Moreover, a ratio of magnitudes of indicated
maxima in total result [with account for Fig.\ 2 (b)] is sensitive to the
relative contribution of Fig.\ 2 (a) and (b) graphs. The relative
magnitude of the maximum at $E_{ex}=0$ decreases with increasing  a
partial yield of continuum [Fig.\ 2 (b)] in the intermediate state. This
fact appears to be rather important for fitting the data.

A rapid variation of the $^4_{\Sigma}$He production amplitude as function
of $E_{ex}$ was not taken into account in the analysis of Ref.\ 1. It
made the procedure not quite correct. Let's look what are the results of
the correct account for the production mechanism, corresponding to the
graphs of Fig.\ 2 (a) and (b). We shall begin from attempt to describe
$E_{ex}$ spectrum with the parameters from Ref.\ [1], that is, the binding
energy 4.4 MeV (it corresponds to $E_{ex}=-7$ MeV) and the width 7 MeV.
The best description for this case is shown in Fig.\ 4 (a). It is necessary
to accept here that the ratio of Fig.\ 4 (a) and (b) contributions is not
more than 1:5. Otherwise there would be too large enhancement at $E_{ex}=0$
in obvious contradiction with the data.\footnote{We mean the ratio without
account of Breit--Wigner resonance factor. The contribution of Fig.\ 2 (b)
graph to actual spectrum remains small as the resonance factor hardly
suppresses the whole area $E_{ex}>10$ MeV.} It is possible to see that the
peak is described not so well, especially the left wing.

The situation can be improved by a modification of $^4_{\Sigma}$He
parameters. Various versions of the fitting procedure have shown that the
best description of $E_{ex}$ spectrum could be obtained with the binding
energy 5.4 MeV (it corresponds to $E_{ex}=-8$ MeV) and the width 8 MeV.
This fit is shown in Fig.\ 4 (b). The smaller width would lead to a poor
description for the left wing of the resonance peak. The larger width would
lead to too strong peak at $E_{ex}=0$. The latter is also essential in
another respect. As indicated above, the technique for inclusion of
$\Lambda$ production tail is incomplete. If the solid curve in Fig.\ 1 were
more rapidly decreasing, then, after its subtraction, the resonance left
wing would be broader. It would demand the larger value of the width.
However, as it appears, the width more than $8.0 \div 8.5$ MeV is forbidden
as it would too strengthen the peak near $E_{ex}=0$. Besides, to keep the
magnitude of this peak in reasonable limits, it is necessary to suppose
that the contribution of multi--particle intermediate states in Fig.\ 2 is
several times more than the contribution of two--particle states. From here
follows that the probabilities of virtual $^4_{\Sigma}$He decays to three
and four--particle channels are much larger than to two--particle ones.

In summary we shall mark the following: \\
1. The amplitude of $^4_{\Sigma}$He production was shown to be a sharply
varying function of mass just in the resonance region. \\
2. It results in a small, but noticeable shift of $^4_{\Sigma}$He
parameters in comparison with the results of Ref.\ [1]. The central values
of both the binding energy and the width are increased on 1 MeV. This
shift, though is not large and does not exceed the limits of the errors
indicated in Ref.\ [1], nevertheless can be important for estimations of
$\Sigma$--nuclear interaction [10].\\
3. From the comparison of the cross section near $E_{ex}=0$ with
calculations, the additional evidence is obtained that $^4_{\Sigma}$He
width does not exceed $8.0 \div 8.5$ MeV. The indication is also obtained
on preferred role of multiparticle channels for virtual $^4_{\Sigma}$He
decay.\\
4. The considered case can be of interest in more general aspect as the
unique example of a resonance on a sharply varying background.\\
5. The appearance of more statistically based data will require to refine
the calculations in several points: (a) using a realistic $^4$He wave
function; (b) account of a form factor in the vertex of the resonance
production in Fig.\ 2 (a); (c) elaboration of a reliable model for the
$\Lambda$ production tail in $({\mbox K}^-,{\pi}^-)$ processes.

{\it Acknowledgments}: The author is indebted to O. D. Dalkarov and
T. E. O. Ericson for discussions. He also appreciate the hospitality of
The Svedberg Laboratory of the Uppsala University where a part of this
investigation was done.

\newpage
\begin{figure}
\begin{picture}(400,400)
\put(40,20){\epsfig{file=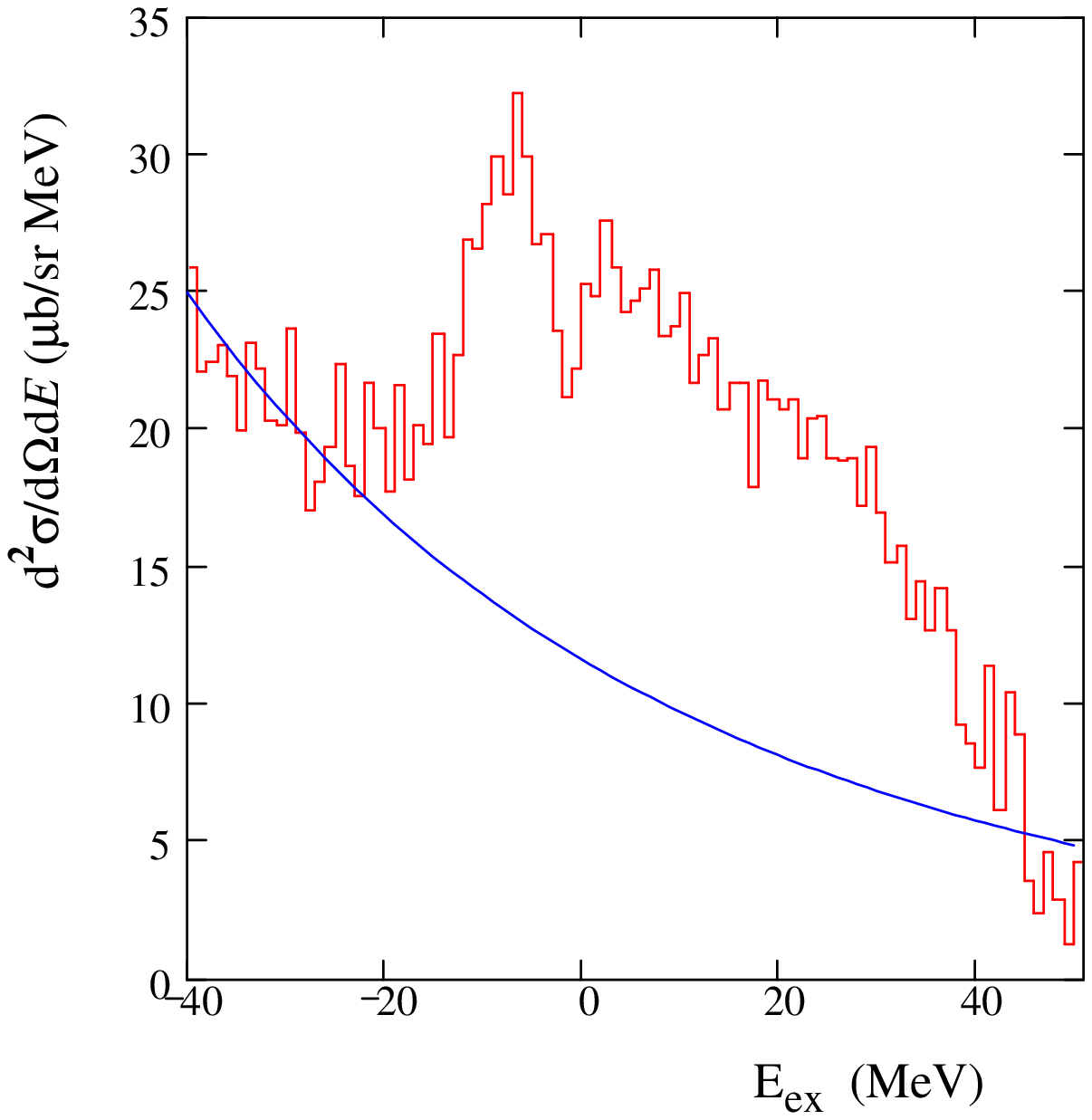, bbllx=94pt, bblly=335pt,
bburx=441pt, bbury=689pt}}
\end{picture}
\caption{The data of Ref.\ [1] on the differential cross sections of
the reaction $^4\mbox{He}(\mbox{K}^-,{\pi}^-)$. The solid curve is
the approximation of the tail of direct $\Lambda$ production.}
\end{figure}

\begin{figure}
\begin{picture}(200,600)
\put(110,20){\epsfig{file=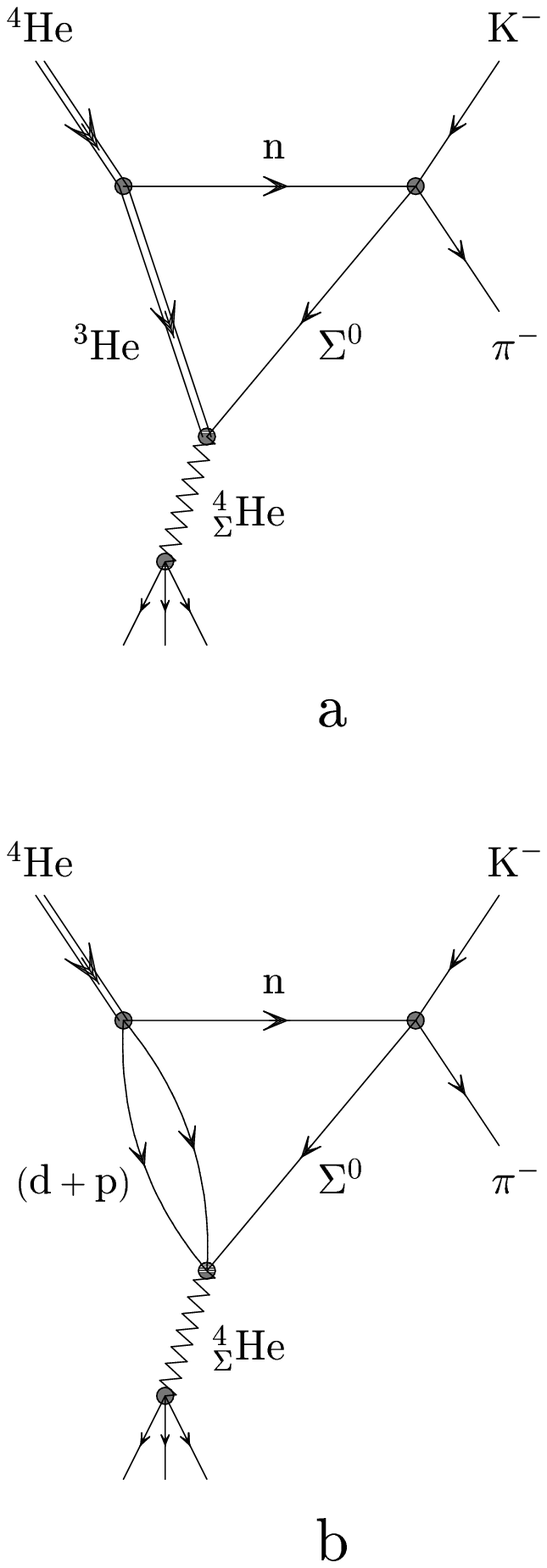, bbllx=196pt, bblly=164pt,
bburx=385pt, bbury=696pt}}
\end{picture}
\caption{Graphs for $^4_{\Sigma}$He production in the reaction
$^4\mbox{He}(\mbox{K}^-,{\pi}^-)$.}
\end{figure}

\begin{figure}
\begin{picture}(450,340)
\put(50,20){\epsfig{file=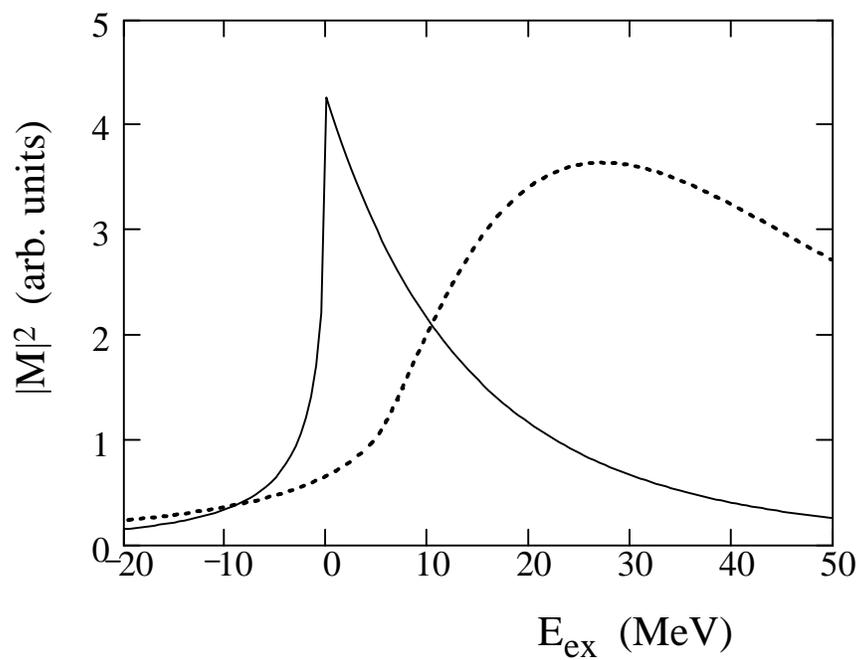, bbllx=109pt, bblly=330pt,
bburx=431pt, bbury=578pt}}
\end{picture}
\caption{$|M_{\triangle}|^2$ for the triangle graphs of Fig.\ 2 with
two--particle (solid curve) and three--particle (dotted curve)
intermediate states.}
\end{figure}

\begin{figure}
\begin{picture}(300,580)
\put(60,20){\epsfig{file=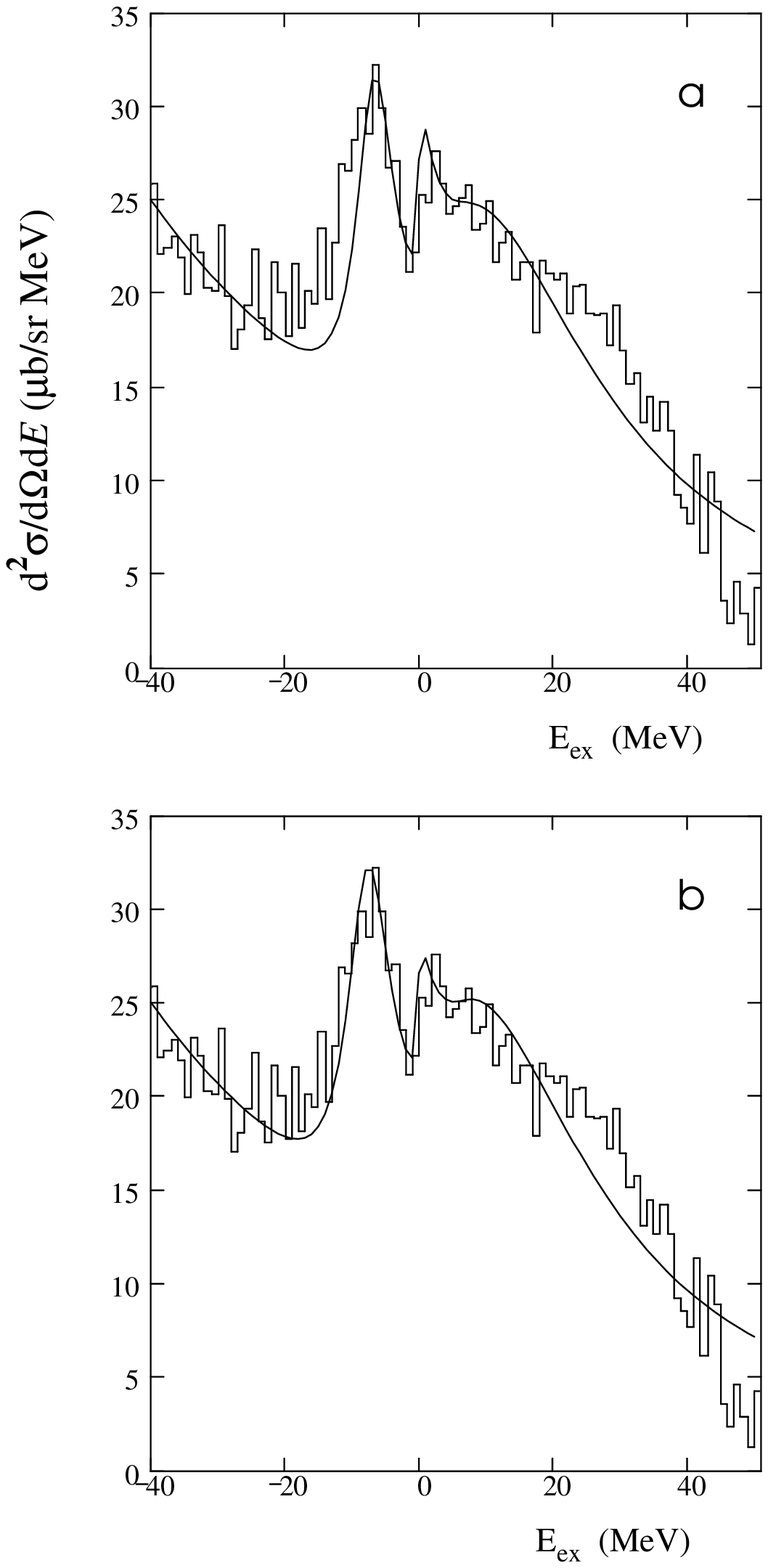, bbllx=134pt, bblly=87pt,
bburx=440pt, bbury=708pt, height=19cm}}
\end{picture}
\caption{Theoretical description of $E_{ex}$ spectrum for
the reaction $^4\mbox{He}(\mbox{K}^-,{\pi}^-)$: (a) with $^4_{\Sigma}$He
parameters from Ref.\ [1]; (b) with the binding energy 5.4 MeV and
the width 8 MeV.}
\end{figure}

\end{document}